# EXPERIMENTAL STUDY OF BEAM DYNAMICS IN THE PIP-II MEBT PROTOTYPE*

A. Shemyakin†, J.-P. Carneiro, B. Hanna, V. Lebedev, L. Prost, A. Saini, V. Scarpine, Fermilab, Batavia, IL 60510, USA
V.L.S. Sista, Bhabha Atomic Research Centre (BARC), Mumbai, India
C. Richard, Michigan State University, East Lansing, MI, USA

*Abstract*

The Proton Improvement Plan, Stage Two (PIP-II) [1] is a program of upgrades proposed for the Fermilab injection complex, which central part is an 800 MeV, 2 mA CW SRF linac. A prototype of the PIP-II linac front end called PIP-II Injector Test (PIP2IT) is being built at Fermilab. As of now, a 15 mA DC, 30-keV H- ion source, a 2 m-long Low Energy Beam Transport (LEBT), a 2.1 MeV CW RFQ, followed by a 10 m Medium Energy Beam Transport (MEBT) have been assembled and commissioned. The MEBT bunch-by-bunch chopping system and the requirement of a low uncontrolled beam loss put stringent limitations on the beam envelope and its variation. Measurements of transverse and longitudinal beam dynamics in the MEBT were performed in the range of 1-10 mA of the RFQ beam current. Almost all measurements are made with 10 μs beam pulses in order to avoid damage to the beam line. This report presents measurements of the transverse optics with differential trajectories, reconstruction of the beam envelope with scrapers and an Allison emittance scanner, as well as bunch length measurements with a Fast Faraday Cup.

## PIP2IT WARM FRONT END

The PIP2IT warm front end (Fig. 1) has been installed in its nearly final configuration [2].

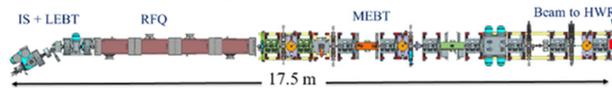

Figure 1: PIP2IT warm front end (top view).

The combination of the ion source and LEBT can deliver up to 10 mA at 30 keV to the RFQ with pulse lengths ranging from 1 μs to 16 ms at up to 60 Hz, or a completely DC beam. An atypical LEBT transport scheme [3] minimizes changes of the beam properties throughout a pulse due to neutralization, which allows to tune the beam line at a short pulse length (typically 10 μs). Following the RFQ is a long MEBT, which provides transverse and longitudinal focusing to match the 2.1 MeV beam into the Half-Wave Resonator (HWR) cryomodule. As the latter is not yet installed, the beam line currently ends with a high-power dump capable of dissipating 10-20 kW, depending on the beam size.

## PIP2IT MEBT

The present MEBT configuration is shown in Fig. 2. The MEBT transverse focusing is provided by quadrupoles [4]. referred as either F or D type according to their yoke length, 100 or 50 mm, which can be powered to focus either in horizontal (+) or vertical (-) directions. The quadrupoles are grouped into two doublets followed by seven triplets, where the magnets are arranged as $F^-$-$F^+$ and $D^-$-$F^+$-$D^-$, respectively. The spaces between the focusing groups are addressed as "sections" (650-mm long flange-to-flange for sections #1 through #7, and 480 mm for section #0). Each group includes a Beam Position Monitor (BPM), whose capacitive pickup is bolted to the poles of one of the quadrupoles and is followed by an assembly with two (X/Y) dipole correctors. The distance between centers of the triplets is 1175 mm.

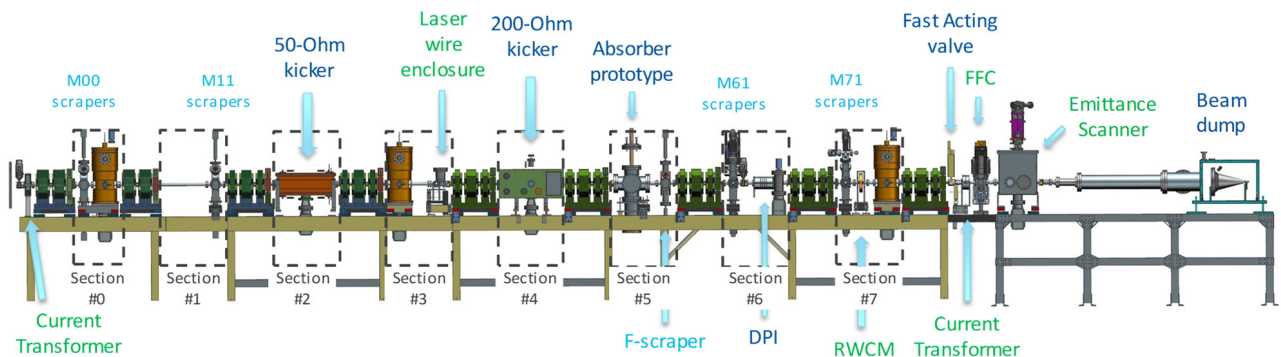

Figure 2: Medium Energy Beam Transport line (side view).

The prototype kickers [5] are installed in sections 2 and 4, and the Differential Pumping Insert (DPI) is in section 6. The 200 mm (L) × 10 mm (ID) beam pipe of the DPI as well as the 13-mm high gaps in the protection electrodes, placed on both sides of each kicker, are the aperture limitations in the MEBT. Otherwise, the typical vacuum pipe ID is 30 mm.

Longitudinal focusing is provided by 3 bunching cavities in sections 0, 3, and 7, which, if phased for acceleration, can increase the beam energy by up to 100 keV each. Movable scrapers [6] installed with the main goal to protect the cryomodules against an errant beam or halo were also used to measure the beam size. Shown in Fig. 2 are 4 sets of 4 scrapers (each set consists of a bottom, top, right and left scraper) plus a temporary set of two scrapers (a.k.a. F-scraper, top and right).

Current transformers are located at the beginning and end of the MEBT. An emittance scanner and Fast Faraday Cup (FFC) (moved to various locations) were used to characterize the beam emittance. A Resistive Wall Current Monitor (RWCM) completes the set of diagnostics available.

## TRANSVERSE OPTICS

Reconstruction of the beam transverse optics in the MEBT was performed in two steps [7]. First, the beam dipole motion was characterized using differential trajectories analyses, and the calibration of magnetic elements was adjusted in the optics model to fit the measurements. Then, the measurements of the transverse beam size along the MEBT were used to reconstruct the Twiss functions of the beam coming out of the RFQ and simulate the beam envelope in the line. This knowledge allows to adjust the beam position and size in a predictable manner.

### Differential Trajectory Analysis

A Java program developed for the differential trajectory measurements at the PIP2IT, records the BPM positions with the nominal settings and then when one of the dipole correctors is changed. The difference in the BPM readings is then compared to the optical model in OptiM [8]. The procedure is repeated for all available correctors, and for each case, the calibrations of the correctors and quadrupoles are adjusted in OptiM, with respect to initial values based on magnetic measurements, to match the data points. After several series of measurements and adjustments, the procedure converges and the model fits well the differential trajectories results in all cases, like the ones shown in Fig. 3.

Typically, the quadrupole calibrations obtained from beam measurements are consistently lower by 5-10% than those found from the magnetic measurements. Note that all magnets were measured at BARC (where they were manufactured) and several were re-measured at Fermilab, where the results from BARC were reproduced well, with difference in calibrations < 1%. The accuracy of the beam measurements, determined by the beam jitter (see below) and drifts, is estimated to be ~4% and cannot explain the deviation. While the discrepancy has not been resolved, in the following analysis we use the calibrations from the beam measurements.

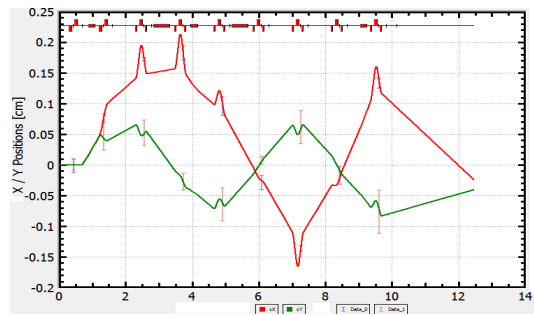

Figure 3: Response of trajectories to changing the current in the first dipole correctors by 0.4 A: horizontal $x$ (red) and vertical $y$ (green). The measured data points are averages of 50 pulses, and the error bars are the rms scatter. The solid lines are the corresponding simulations.

### Beam Size Measurements

Transverse beam sizes along the MEBT are measured primarily with scraper scans fitted to an integrated Gaussian distribution (see more details in [9]). In addition, the Allison scanner, installed at the end of the beamline, provides the vertical phase space portrait hence the vertical beam size.

To describe the beam envelope along the MEBT, the initial transverse Twiss parameters at the exit of the RFQ were defined through an iterative process using TRACEWIN [10] simulations to fit to the measured rms beam sizes at the first three scrapers. Figure 4 shows the reconstructed transverse rms envelope, which agrees with all measured sizes within their typical reproducibility of ~10%.

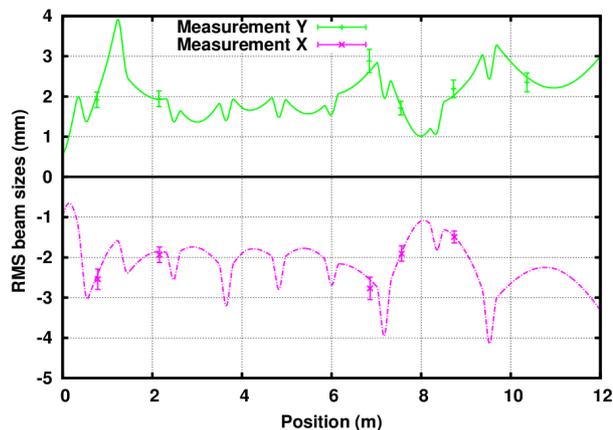

Figure 4: Rms beam envelope along the MEBT simulated with TRACEWIN. Horizontal envelope is shown negated for the presentation purpose. Error bars are +/- 10% of the measured sizes. Normalized rms transverse and longitudinal emittances were assumed to be 0.2 μm and 0.28 μm, respectively. Beam current is 5 mA.

### Beam Tuning

The accuracy of the optics model helped with beam tuning. The beam envelope presented in Fig. 4 is optimized for the MEBT line operating with two kickers and the DPI. Consequently, the vertical beam size is lower than the horizontal in the kickers, and both sizes are small when passing through the 10-mm aperture of the DPI. In high-power runs, the beam size in the dump was increased by over-focusing the beam with the last triplet.

Also, the model easily predicts how to combine setting changes to several correctors to move the beam in specific locations without disturbing the trajectory elsewhere.

*Beam Jitter*

The beam in the MEBT experiences a significant pulse-to-pulse jitter that affects the accuracy of the measurements, the effective emittance, and aperture limitations. The amplitude varies dramatically along the beam line, reaching up to 0.2 mm rms. In an attempt to localize the source of that jitter, the BPM readings of 10 µs x 20 Hz pulses were recorded over 35 minutes from all BPMs. The resulting matrix was analysed with Singular Value Decomposition similar to Ref. [11]. The analysis showed that the noise is dominated by a single spatial component, which eigenvalue exceeded the next closest one by a factor of ~10 (Fig. 5). Components beyond the second one are already at the noise floor. The FFT analysis of the noise temporal structure showed that the noise is dominated by low frequencies,

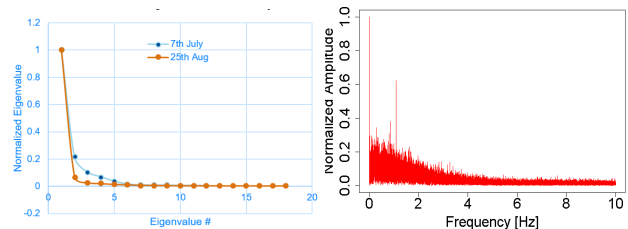

Figure 5: Characteristics of the BPMs noise. Left-eigenvalues. The data set marked "7th July" represents 10 Hz x 10 min set. Right – FFT of the signal in one of the BPMs.

Comparing the first two spatial eigenvectors (Fig. 6) to the MEBT betatron modes clearly indicates that the beam jitter originates upstream of the MEBT.

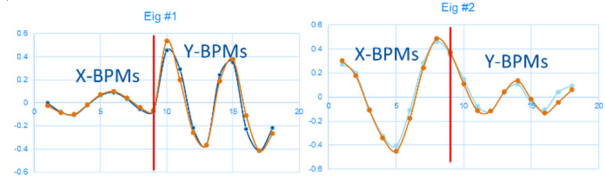

Figure 6: Comparison of the first two eigenvectors (orange) with betatron modes (blue).

After subtracting the contribution of the first two modes from the original data, the remaining noise is at the level of several µm, probably defined by the electronics.

To further localize the noise source, we varied the last two (out of 3) LEBT solenoids and found two combinations of their currents (in addition to the nominal settings) for which the RFQ transmission was good (>95%). The solenoids rotate the beam as well as the plane of the jitter proportionally to the sum of Amp-turns in them, while the motion in the RFQ and MEBT is uncoupled. The BPM signals were recorded for both cases, and the same SVD analysis was performed. The plane of oscillation of the first spatial eigenvector was found changing in agreement with angles expected from the rotation by the solenoids, indicating that the source of the jitter is upstream of the second solenoid. Unfortunately, as of now, the jitter has not been eliminated. The present speculation is that it comes from within the ion source where faint indications of the presence of the same 1.09 Hz line as in Fig. 5 were observed.

*Road to Beam Tails Analysis*

So far, the beam properties were characterized in terms of either the centroid motion or rms sizes. To analyse the beam transverse tails, we initially intended to use the scraping system. It was not successful. On one hand, the noise of the current measuring devices is too large to resolve variations below 1% when a scraper is moved into the beam. On the other hand, the signals from the scrapers themselves drop to nearly zero if their plates are at the ground potential because of secondary electron emission. Since the scraper currents are to be included into the Machine Protection System (MPS), the plates are biased by +100 V. In this case the scraper current starts already rising when the scraper plate is far from the beam. When the plate is deep inside the beam, the scraper reading is typically 10-20% higher than the intercepted beam current and fluctuates significantly. We interpret this as an indication of the presence of a significant amount of secondary electrons in the vacuum pipe, e.g. originated by lost or reflected ions from the pipe's walls. While such behaviour is tolerable for the purpose of the MPS, it does not allow measuring the tails of the particles distribution. We hope that the implementation of a negatively biased wire scanner will provide much more consistent readings.

Another available tool is the MEBT Allison scanner. A dedicated Python application is being written to better analyze the tails. First, attention is paid to the background analysis. In the present LabView program, inherited from SNS, the background is rejected at the level equal to 1% of the maximum signal, distorting the output for low beam currents or large footprints. Instead, the new code analyses the level of the background noise far from the beam and reject the background at the level of several times the rms noise (Fig. 7). The typical ratio of the maximum signal to the cut-off is 0.5%.

Then, efforts are being made to define more consistently the beam core. The rms definition of the Twiss parameters depends on the level of the noise cut since the tails are generally phase-dependent. In addition, the Twiss parameters change if tails are cut by a scraper, making the analysis of the scraping efficiency difficult in terms of maximum action. We are implementing a procedure that calculates the Twiss parameters for only 50% of the total phase portrait integral, composed by the pixels with highest intensities. One of the consequences of using such "central" definition is that in the core the pixels corresponding to the same action have the same intensity, i.e. the core distribution is independent on the phase (Fig. 7). Note that this "central" definition is likely more consistent with the envelopes derived from scraper measurements since fitting to a Gaussian distribution essentially ignores the behaviour of the tails.

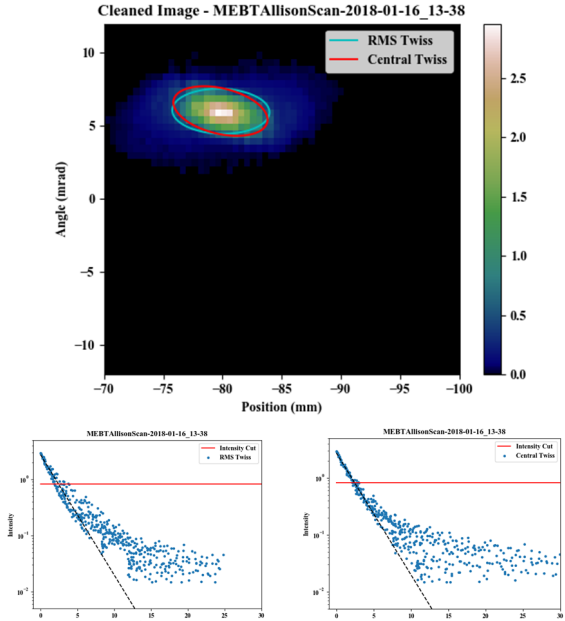

Figure 7: Illustration of the different Twiss parameters calculation. Top- the image cleaned with the Python code. Bottom- pixel intensity (in mV) vs the action (in µm) for the rms (left) and "central" definitions. In the core, the scatter of intensities is significantly lower in the right plot.

## LONGITUDINAL MOTION

Longitudinal focusing at PIP2IT is provided by three bunching cavities. Normally the cavities' phases are set to -90º with respect to the beam, i.e. the bunches are longitudinally focused without changing the average energy of the ions.

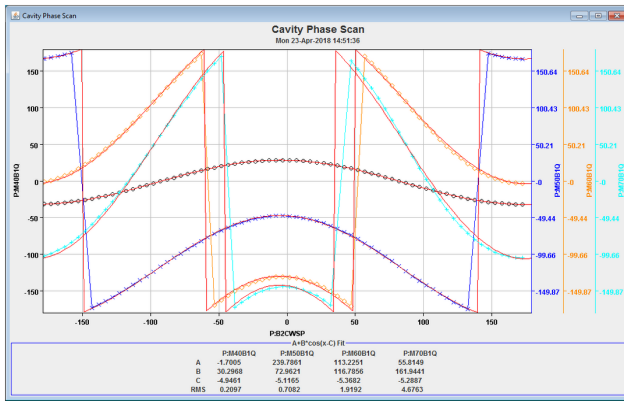

Figure 8: Screenshot of the cavity phasing program showing the dependence of 4 BPM phases on the reference phase of bunching cavity #2. Cavity voltage is 60 kV. In this specific case, the cavity phase offset needs to be adjusted by 5º.

The phasing is made by rotating the cavity reference phase by 360º, recording the resulting changes in phases of the downstream BPMs and fitting them to sinusoids (Fig. 8). The found phase offset is then corrected via the LLRF settings so that the zero degrees of the reference phase corresponds to maximum acceleration. Typical scatter in these measurements is ~0.5º.

Note that the cavities' voltages can also be deduced from phasing measurements (as in Fig. 8) since the amplitude of BPM phase variations is proportional to the cavity voltage. These measurements provide amplitudes that are ~10% higher than previously established calibrations. The reason of this discrepancy is under investigation.

The longitudinal charge distribution of the bunches is measured with a Fast Faraday Cup, which can be moved vertically in and out of the beam path. Its 0.8-mm entrance hole in the ground electrode cuts a beamlet, which current is measured by a collector. The 1.7 mm gap between the ground electrode and collector results in widening of the measured signal in comparison with the actual bunch length. According to estimations in [12], a point charge flying at 20 mm/ns (equivalent to 2.1 MeV) would generate a pulse with an rms width of 25 ps. For all bunch length measurements at PIP2IT, this correction is negligible. All FFC measurements are made with 10 µs x 1 Hz pulsing.

The bunch length measurements were carried out in two locations, first in section 6 and then at the end of the beam line as shown in Fig. 2. The measured distributions are characterized by the rms bunch length and integral, which are obtained from fitting the signals to a Gaussian distribution. A typical FFC signal and corresponding Gaussian fit are shown in Fig. 9.

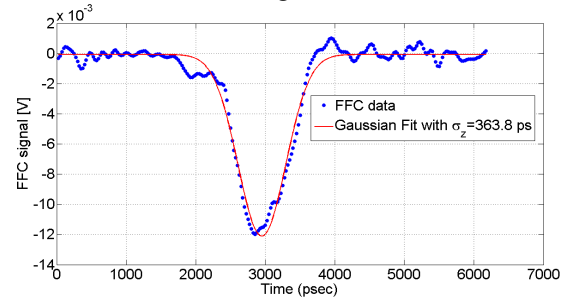

Figure 9. One period of the FFC signal (blue) and its Gaussian fit (red). The FFC is at the end of the beam line. Beam current is 9.3 mA.

Location of the FFC in section 6 is optimum for reconstructing the bunch longitudinal emittance since the dependence of the bunch length on the voltage of bunching cavity #2 upstream exhibits a minimum (Fig. 10).

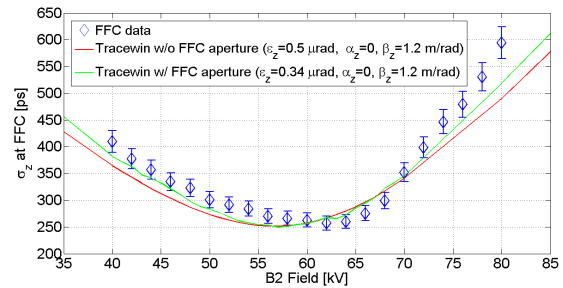

Figure 10: Rms bunch length vs voltage of bunching cavity #2 (blue) and two fitting curves (see text).

The initial interpretation of the data yielded a large longitudinal emittance, 0.5 µm rms normalized (red curve in

Fig. 10), in a strong disagreement with simulations. The contradiction was traced to the implicit assumption that the bunch length measured at the beam center is representative of the entire beam, which is valid for a fully uncoupled particle distribution. Detailed measurements of the bunch length in various positions across the beam clearly showed that this assumption is incorrect: the bunch length is consistently lower toward the beam edges (Fig. 11).

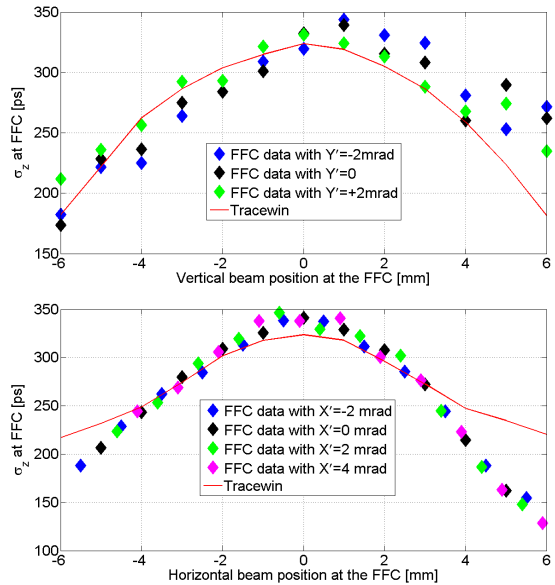

Figure 11. Rms bunch length vs vertical (top) and horizontal (bottom) position across the beam. Different colors of data points correspond to different angles between the beam and the FFC axis. The red solid curves are simulations with TraceWin assuming an uncoupled 6D distribution at the exit of the RFQ. Beam current is 5 mA. The beam transverse rms size is 2.6 mm in X and 2.0 mm in Y.

For a more adequate comparison with measurements, the bunch length of a beamlet cut by same 0.8 mm aperture was simulated by TraceWin. In this case, the simulation exhibits the behaviour observed experimentally (solid curves in Fig. 11), and the best fit for the bunch longitudinal emittance, 0.34 μm, is close to the one expected from RFQ simulations (green curve in Fig. 10). These simulations use as an input file a fully uncoupled 6D Gaussian distribution, primarily because the procedure of adjusting the initial Twiss functions is clear in this case. The significant difference between the rms bunch length of the entire beam and the central beamlet appears only with significant space charge and only after propagation through a large part of the MEBT. At the same time, the beam longitudinal emittance stayed essentially constant along the MEBT.

Note that analyses of the distribution coming out of the RFQ in simulations with TOUTATIS [10] showed a difference between the "central" and overall bunch length comparable to numbers in Fig. 11. Ignoring this effect in the present MEBT simulations may be affecting the accuracy of comparison with measurements.

## SUMMARY

The transverse optics of the PIP2IT MEBT was reconstructed by first adjusting calibrations of the magnets based on analyzing the beam dipole motion and then defining the initial Twiss parameters through fitting the beam sizes measured along the beam line. Good understanding of the optics helps with tuning the beam, e.g. allowing to predict the beam envelope within 10% or look for the source of the beam position jitter.

The longitudinal optics is defined with less certainty, though its study follows the same combination of analyzing the dipole motion by BPM phases and the rms bunch length by the FFC. One of the results of the latter measurements is that there is significant coupling between the transverse and longitudinal beam distributions.

## ACKNOWLEDGMENT

The authors are thankful to the many people who built the PIP2IT MEBT and helped with the measurements, including R. Andrews, C. Baffes, B. Chase, A. Chen, E. Cullerton, N. Eddy, J. Einstein-Curtis, B. Fellenz, B. Hartsell, M. Kucera, D. Lambert, R. Neswold, D. Peterson, A. Saewert. FFC RF design by D. Sun was crucial for the bunch measurements. Scanning programs were written by W. Marsh. Support and encouragement from P. Derwent is highly appreciated.